# The effect of confinement and defects on the thermal stability of skyrmions

Valery M. Uzdin[1,2]; Maria N. Potkina[1], Igor S. Lobanov[2], Pavel F. Bessarab[2,3], Hannes Jónsson[3]

[1]*Physics Department, St.Petersburg University, St.Petersburg, 198504, RUSSIA*
[2]*University ITMO, St. Petersburg 197101, RUSSIA*
[3] *Faculty of Physical Sciences, University of Iceland, 107 Reykjavík, ICELAND*

**Abstract**

The stability of magnetic skyrmions against thermal fluctuations and external perturbations is investigated within the framework of harmonic transition state theory for magnetic degrees of freedom. The influence of confined geometry and atomic scale non-magnetic defects on the skyrmion lifetime is estimated. It is shown that a skyrmion on a track has lower activation energy for annihilation and higher energy for nucleation if the size of the skyrmion is comparable with the width of the track. Two mechanisms of skyrmion annihilation are considered: inside the track and escape through the boundary. For both mechanisms, the dependence of activation energy on the track width is calculated. Non-magnetic defects are found to localize skyrmions in their neighborhood and strongly decrease the activation energy for creation and annihilation. This is in agreement with experimental measurements that have found nucleation of skyrmions in presence of spin-polarized current preferably occurring near structural defects.

**Introduction**

Magnetic skyrmions could form the basis for ultra-dense and ultrafast magnetic memory, because of their small size, high mobility and topological stability. To explore their applicability as memory elements, an assessment of the stability of skyrmions is important, in particular when their intrinsic size is comparable to the distance separating them or the size of the domain in which they reside [1-3]. An important issue is also the interaction of skyrmions with structural defects and impurities, which will inevitably be present. Such defects could have detrimental effects and shorten the skyrmion lifetime. On the other hand, controlled introduction of defects could possibly also be used in a beneficial way in memory devices.

Defects in the magnetic and crystal structure, such as nonmagnetic atoms, regions with altered anisotropy, or magnetic atoms of a different type, can significantly affect the structure of the skyrmions, their dynamics and stability [1]. Experiments on the creation and destruction of skyrmions induced by an electrical current from the tip of a tunneling microscope in external magnetic field have shown that skyrmions are usually formed near structural defects [4]. It is also known from experiments [5] as well as from the modeling on the basis of the Landau-Lifshitz equation [6] that the skyrmions can be localized near the defects and in this case additional effort is required to dissociate them from the defects. Therefore, it is of interest to analyze the activation energy for skyrmions in the presence of defects, and also the joint effect of defects and finite sample size.

The stability of magnetic skyrmions in confined geometry and in the presence of atomic-scale, non-magnetic defects is studied here within the framework of transition state theory for magnetic degrees of freedom [7]. The energy surface of the system is described with a Heisenberg-like Hamiltonian including exchange interaction between nearest neighbor spins, anisotropy energy, Dzyaloshinskii-Moriya interaction, and Zeeman energy representing interaction with an external magnetic field [8]. The calculations are carried out on a two-dimensional triangular lattice with periodic boundary conditions along one axis and free boundaries in the perpendicular direction. This setup makes it possible to study the stability of a skyrmion on a "track" as a function of the width of the track.

The multidimensional energy surface that specifies the energy of the system as a function of the angles defining the direction of all the magnetic moments is analyzed to identify local minima corresponding to the skyrmion and homogeneous ferromagnetic (FM) states as well as the minimum energy path (MEP) connecting the minima. The maximum along the path is a first order saddle point on the energy surface and gives the activation energy for transitions between the states. The MEP is the transition path with maximum statistical weight and shows the most probable mechanism for the transition [7-9].

Two possible mechanisms of skyrmion annihilation and nucleation are considered: Inside the track and through the boundary of the track. For each mechanism, the activation energy is calculated as a function of the width of the track. In addition, the effect of nonmagnetic defects on the stability of the skyrmion is investigated. The results of these calculations are consistent with experimental observations showing skyrmions located at defects [5] and the formation of skyrmions near defects when a spin polarized current is applied through the tip of a microscope [4].

**Model**

The energy surface of the system is described with a Heisenberg type Hamiltonian

$$H = -J\sum_{<i,j>}\mathbf{S}_i\mathbf{S}_j - \sum_{<i,j>}\mathbf{D}_{ij}[\mathbf{S}_i\times\mathbf{S}_j] - \mu\sum_i\mathbf{B}\,\mathbf{S}_i - K\sum S_{i,z}^2 \qquad (1)$$

Here $J$ is the exchange parameter for nearest neighbor magnetic moments, $\mathbf{D}_{ij}$ is the Dzyaloshinsky-Moriya vector lying in the plane of the sample perpendicular to the vector connecting atomic sites $i$ and $j$, K is the anisotropy constant, B is the external magnetic field and $\mathbf{S}_i$ is a three-dimensional vector of unit length in the direction of the magnetic moment at site $i$ of a two-dimensional triangular lattice. The magnitude of the moment, $\mu$, is assumed to be the same for all sites, $\mu = 3\ \mu_B$. The summation $<i, j>$ in (1) runs over all pairs of nearest neighbor sites. The numerical values of the parameters are taken from [10] and correspond to experimentally observed skyrmions in the Pd/Fe/Ir(111) system [4], $\mu B = 0.093J$, $K = 0.07J$, $D_{ij} = 0.32\ J$, $J=7$ meV.

The MEPs were calculated using the geodesic nudged elastic band method (GNEB) [9].

**Results**

A. Stability of a skyrmion on a track

Fig. 1 shows the calculated skyrmion configurations corresponding to local minima on the energy surface for a track of width $D = 70, 25$, and $14$ atomic rows. Periodic boundary conditions are used along the horizontal direction but free boundaries in the orthogonal direction.

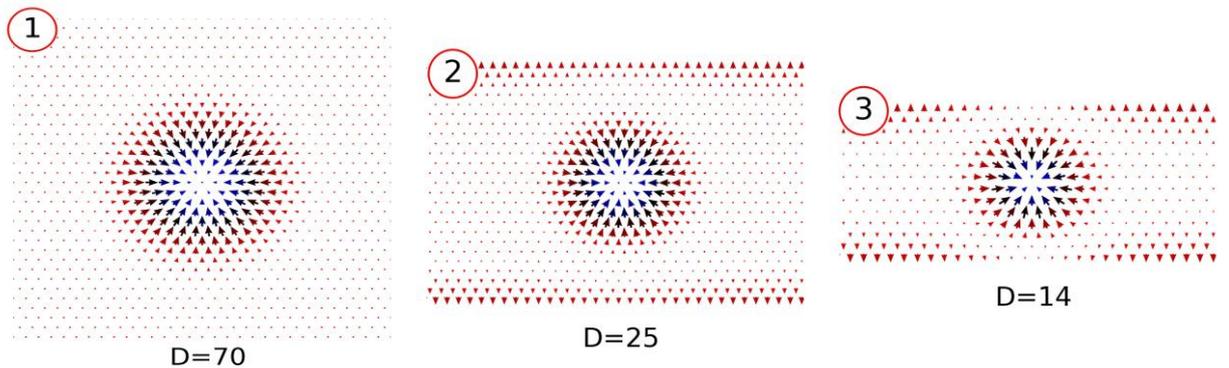

Fig. 1 Skyrmion on tracks of different width, *D*: (1) *D*=70, (2) *D* =25, (3) *D* =14.

The size of the skyrmion decreases as the width of the track becomes smaller. At the boundaries of the track the magnetic moments deviate from the direction orthogonal to the surface due to the uncompensated Dzyaloshinskii-Moriya interaction. When the intrinsic size of the skyrmion approaches the width of the track, the skyrmion structure at the boundary is significantly distorted and this raises the total energy of the system.

The MEPs for transitions between skyrmion and FM states in these systems are presented in Fig. 2. Two mechanisms of skyrmion annihilation were considered: Inside the track and through the boundary. MEPs for the first mechanism are shown in Fig. 2a for the three values of the track width. As the width, D, decreases, the skyrmion energy increases with respect to the energy of the

FM state. The energy at the saddle point also increases but less, since the size of the noncollinear structure at the saddle point is smaller (see inset in Fig. 2a). As a result, the activation energy for skyrmion annihilation is smaller in a narrow track. Correspondingly, the activation energy for skyrmion nucleation is larger. An estimate of the pre-exponential factor in the Arrhenius rate law within harmonic transition state theory [7] gives a value on the order $10^{12}$, varying slightly with the width of the track.

MEPs for the annihilation of skyrmion at the track boundary are shown in Fig. 2b. Here, the variation of the activation energy with the width is similar to annihilation inside the track, but is somewhat smaller. When the width of the track significantly exceeds the intrinsic size of the skyrmion, the process of annihilation across the boundary can be separated into two stages: displacement of the skyrmion towards the boundary, involving only a small increase in energy, and then annihilation at the boundary. This is clearly seen in the case of D = 70 where the MEP has a long horizontal part corresponding to the movement of the skyrmion as a whole without significant change in shape.

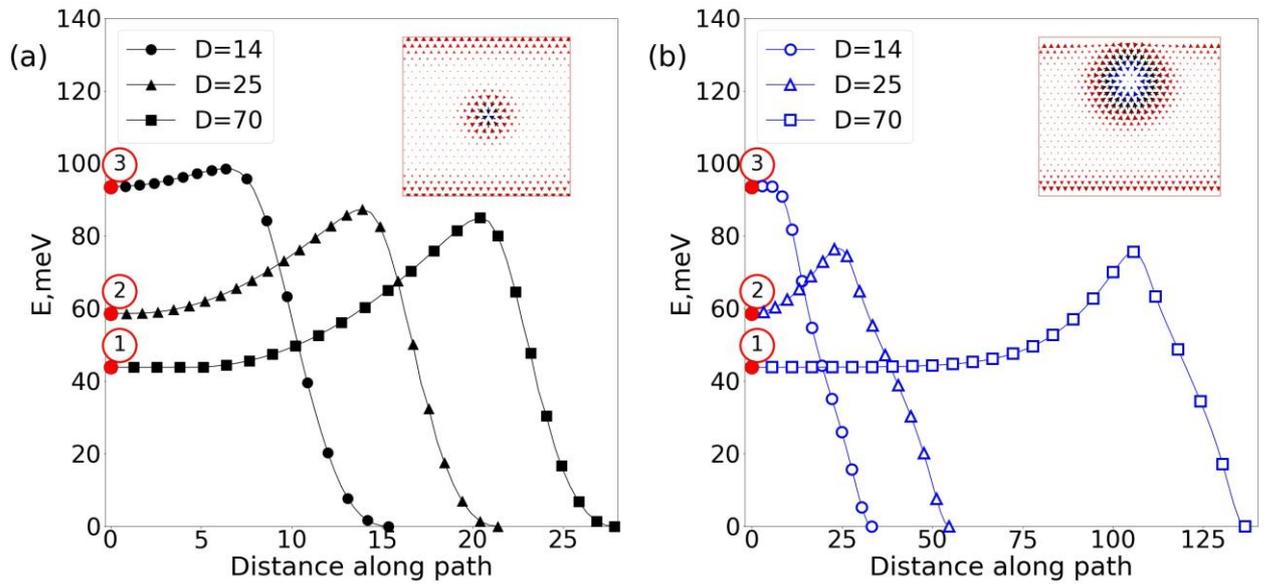

Fig. 2 Minimum energy paths for transitions from skyrmion to FM state for tracks of different width, D, in the interior of the track (a) and at the boundary of the track (b). The number in red circles corresponds to configurations shown in Fig. 1. Insets show the magnetic configuration at the saddle points.

Figure 3 shows the variation of the activation energy for creation and annihilation of the skyrmion inside the track and at the boundary of the track as a function of the track width. For tracks that are wide in comparison with the intrinsic size of the skyrmion, the activation energy does not depend on the width, but annihilation at the boundary has smaller activation energy than annihilation inside the track. As the track is made narrower, the activation energy decreases. For a narrow track, annihilation at the track boundary requires lower activation energy than annihilation inside the track. Activation energy for skyrmion nucleation increases the width of the track is decreased and is also lower for nucleation near the track boundary than inside the track.

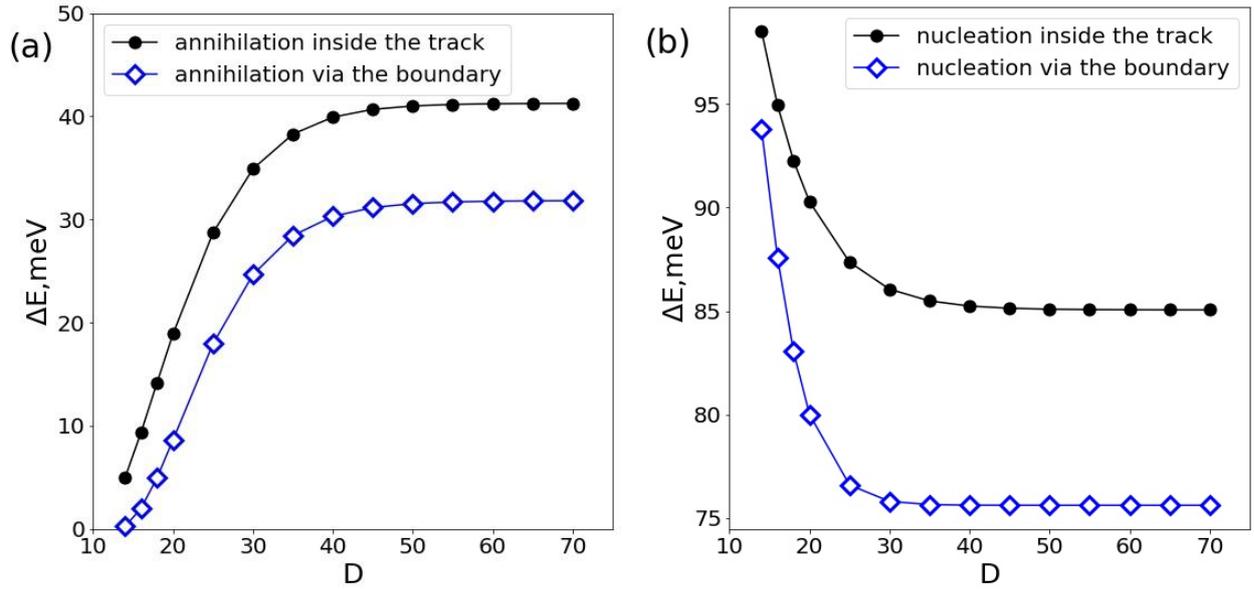

Fig. 3 Variation of the activation energy for annihilation (a) and for nucleation (b) of a skyrmion as the width of the track, D, is varied.

B. Skyrmion localization at non-magnetic defects

In order to investigate the stability of a skyrmion localized at a non-magnetic defect, the corresponding local minimum on the energy surface needs to be found. The calculations show that the minimum energy is achieved when the defect is near the border of the skyrmion, where the magnetic moments has a large in-plane component (see Fig. 4, top spin configuration to the right). This is in agreement with experimental observations [5]. The MEP between this state and the state of skyrmion far from the impurity can then be calculated with the GNEB. This gives the activation energy for pinning and unpinning of a skyrmion to a nonmagnetic defect. The calculations show that the energy of a skyrmion localized at a defect consisting of 3 non-magnetic atoms is lower than the energy of an unpinned skyrmion by 1.7 meV. However, to dissociate from the defect, the skyrmion has to overcome a barrier of 3.3 meV. The existence of an activation energy both for attachment and detachment of the skyrmion to the defect explains results of spin dynamics simulations of skyrmions interacting with a spin polarized current [6] where it was found that for small current densities a previously pinned skyrmion stays pinned whereas an unpinned skyrmion moves around the impurities without getting captured. Since there is an activation energy for the attachment to the defect, a dynamical trajectory at low temperature will not overcome the barrier, but at sufficiently high temperature it might.

Nonmagnetic defects decrease the activation barrier for annihilation as well as for the nucleation of a skyrmion. This is illustrated in Fig. 4 where the activation energy for skyrmion annihilation at a defect consisting of 1 and 3 atoms is shown as function of a track width. For a wide track, the activation energy for skyrmion annihilation near a defect decreases to less than a half as compared with annihilation far from the defect (from 41 meV to 17 meV). This explains why skyrmions usually appear and disappear near defects [5].

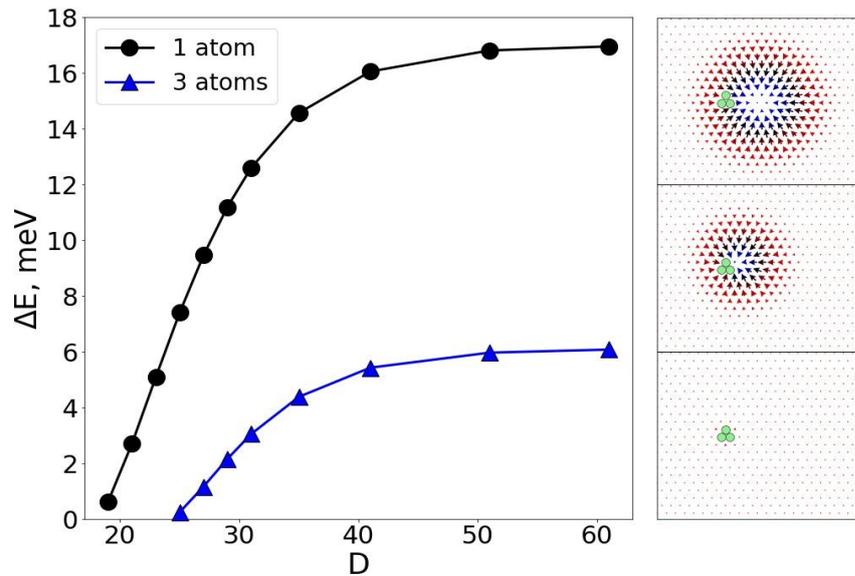

Fig. 4 Left: Activation energy for skyrmion annihilation at defects consisting of 1 or 3 non-magnetic atoms as a function of the track width, D. Right: Configuration of the spins at the initial state (top), saddle point (middle), and final state (bottom).

**Acknowledgments**
This work is supported by the Icelandic Research Fund and the Academy of Finland (grant 278260).